\documentclass[
    ,final
  ]
  {aipproc}

\layoutstyle{8x11single}

\usepackage{bm}
\usepackage{latexsym}

\begin{document}

\title{The two body problem : analytical results in a toy model of relativistic gravity}


\author{Jean-Philippe Bruneton}{
  address={Institut d'Astrophysique de Paris 98 bis bd Arago 75014 PARIS}
}

\begin{abstract}
The two body problem in a scalar theory of gravity is
investigated. We focus on the closest theory to General Relativity
(GR), namely Nordstr\"om's theory of gravity ($1913$). The
gravitational field can be exactly solved for any configuration of
point-particles. We then derive the exact equations of motion of
two inspiraling bodies including the exact self-forces terms. We
prove that there is no innermost circular orbit (ICO) in the exact
theory whereas we find (order-dependent) ICOs if post-Newtonian
(PN) truncations are used. We construct a solution of the two body
problem in an iterative (non-PN) way, which can be viewed as a
series in powers of $(v/c)^{5}$. Besides this rapid convergence,
each order also provides non-perturbative information. Starting
from a circular Newtonian-like orbit, the first iteration already
yields the 4.5 PN radiation reaction. These results not only shed
light on some non-perturbative effects of relativistic gravity,
but may also be useful to test numerical codes.
\end{abstract}

\maketitle

\section{Introduction}
The two-body problem in GR is both an important and difficult
issue. Various approach have been investigated. One generically
needs some assumptions and simulations (numerical relativity) or
perturbatives techniques (PN expansion, e.g \cite{Templatesa3PN}).
In both cases it remains difficult to quantify the errors one
makes. This point is however crucial in order to interpret the
coming-soon data.

It should thus be very interesting to test the numerical codes and
the assumptions usually made in the two-body problem within a toy
model of GR. One may also hope that some generic non-perturbative
effects may be found, that could also occur in GR. This toy model
should be simple enough to treat analytically (at least partially)
the two body problem, but should also be as close as possible to
GR. Metric scalar theories of gravity are perfect candidates of
such toy models.

This idea is not a new one. It has already been used in order to
test numerical codes, see e.g \cite{Gravscalaire1,Gravscalaire2}
or, more recently, the validity of the Quasi Equilibrium (QE)
scheme \cite{GravscalaireQE}. However we wish here to focus on a
particular theory of this class (namely Nordstr\"om's one), which
is, to our mind, the best motivated toy-model of GR, and which has
not been used in the literature. An important point is that
Nordstr\"om's theory respects the Strong Equivalence Principle
(SEP) and thus enables us to consider constant-mass
points-particles, whereas this cannot be justified in others
models considered in the literature. We also want to go further in
the analytical resolution of the two-body problem, by using one of
the great advantages of Nordstr\"om's theory: the linearity of the
field equation.

In Section II we recall the basic features of such metric scalar
theories and notably the particular case of Nordstr\"om's theory.
In Section III, we solve exactly the metric in terms of $N$
arbitrary massive point particles. We then derive the
\textit{exact} equations of motion of such bodies which notably
involve finite self-force terms.

In Section IV, we restrict to the equal-mass two-body case and
investigate the circular motion. We derive the analytical relation
between the radius and the orbital velocity. We compute the exact
energy of the binary in such a circular configuration and show
that the theory does not exhibit an innermost circular orbit
(ICO), whereas such ICOs are found if one works with PN
truncations of the exact energy.

In Section V, starting from this circular solution, we show that
the inspiral motion can be found order by order in power of
$(v/c)^5$, where $v$ is the velocity of the bodies. This behavior
is directly related to the leading multipolar emission of the
system, namely the quadrupolar one. Furthermore, this expansion is
not a PN one since each orders yield non-perturbative information
on the motion.

In Section VI, we show some analytical and numerical results. We
give the exact rate of change of the velocity up to 4.5 PN and the
radius up to 7PN. We also plot these two quantities and by
numerical integration, we plot the corresponding inspiral motion.
This paper is a short version of the self-contained article
\cite{NotreArticle} in which more technical details can be found.
\section{A quick look at Nordstr\"om's theory of gravity}
We first introduce briefly metric scalar theories of gravity. The
action reads
\begin{equation}\label{Action}
\mathcal{S} = -\frac{c^4}{8 \pi G}\int
\sqrt{-\eta}\frac{d^4x}{c}\eta^{\mu\nu}\partial_{\mu}\varphi
\partial_{\nu}\varphi + \mathcal{S}_{\textrm{matter}}[\textrm{matter}
; \tilde{g}_{\mu\nu}\equiv A^2(\varphi)\eta_{\mu\nu}],
\end{equation}
where \(A(\varphi)\) is a given function of the scalar field,
characterizing the theory. The action of matter,
$\mathcal{S}_{\textrm{matter}}$, is a functional of all matter
fields, assumed to be minimally coupled to the ``physical'' metric
$\tilde{g}_{\mu\nu}\equiv A^2(\varphi)\eta_{\mu\nu}$. These
theories thus respect the weak equivalence principle. Throughout
this paper, we will use the sign conventions of \cite{MTW}, and in
particular the mostly-plus signature $-+++$. Throughout the paper
we will use coordinates for which the flat metric $\eta_{\mu\nu}$
takes its fundamental form. We use bold-faced symbols to denote
the three vectors of the Minkowskian geometry.

Since the physical metric is conformally related to the flat one,
there is no coupling of the photon to the scalar field. In
\textit{any} theory of the type (\ref{Action}), there is thus
strictly no light deflection, and all of them are ruled out by
experiment. They anyway share many feature with general
scalar-tensor theories \cite{Def92} and one of them even satisfies
the SEP.

The field equations deriving from action~(\ref{Action}) read
\begin{equation}
\label{Eqinitiales} \Box_{\textrm{flat}}\varphi=-\frac{4 \pi
G}{c^4}A^3(\varphi)A'(\varphi) \tilde{T}
\end{equation}
\begin{equation}
\label{Conservtilde} \tilde{\nabla}_{\mu}\tilde{T}^{\mu\nu}=0,
\end{equation}
where $\tilde{T}^{\mu\nu}$ is the physical stress-energy tensor,
$\tilde{T} \equiv \tilde{g}_{\mu\nu}\tilde{T}^{\mu\nu}$ its trace,
$\tilde{\nabla}_{\mu}$ denotes the covariant derivative with
respect to (w.r.t) the physical metric $ \tilde{g}_{\mu\nu}$, and
$\Box_{\textrm{flat}}$ is the flat d'Alembertian operator. The
scalar curvature may be written as
\begin{equation}
\label{Rtilde2} \tilde{R}=\frac{24 \pi
G}{c^4}A'^2(\varphi)\tilde{T}-6\frac{A''(\varphi)}{A(\varphi)}\tilde{g}^{\mu\nu}\partial_{\mu}\varphi\partial_{\nu}\varphi.
\end{equation}
Therefore, if and only if $A(\varphi)=\varphi$, the theory admits
a purely geometrical and generally covariant formulation
\cite{EinsteinFokker}:
\begin{equation}
\tilde{R}=\frac{24 \pi G}{c^4}\tilde{T}, \,
\tilde{C}^{\lambda}_{\mu\nu\rho}=0, \,
\tilde{\nabla}_{\mu}\tilde{T}^{\mu\nu}=0
\end{equation} where $C$ denotes the Weyl tensor. Using the same reasoning as in
\cite{DamourHouches}, this geometrical formulation suffices to
prove that the SEP is satisfied. Among all others theories of type
(\ref{Action}), this particular one characterized by
$A(\varphi)=\varphi$ is thus the closest one to GR. In the
following, it will be referred to as Nordstr\"om's theory of
gravity \cite{Nordstrom1} (see also
 \cite{Nordstrom2} and the review \cite{ReviewNordstrom}).

To our knowledge, there exist only two gravity theories satisfying
the SEP \cite{Def92}: GR and Nordstr\"om's theory of gravity. The
SEP notably means that the gravitational binding energy of a body
contributes the same to its inertial and gravitational masses, so
that strongly self-gravitating bodies fall in the same way as test
masses in an external gravitational field (up to self-force
effects which can play a major role in the dynamics). This fact
therefore allows us to describe massive bodies as constant-mass
point particles without worrying about their internal structure.
In all other theories of type (\ref{Action}), with a non-linear
matter-scalar coupling function $A(\varphi)$, violations of the
universality of free-fall of self gravitating bodies would appear
already at the first PN level.

In the following, we will be interested in matter consisting of
massive point particles of coordinates $z_A^\mu$, whose Lagrangian
read $ L= -\sum_A m_A c A(\varphi) ds$ where
$ds^2=\eta_{\mu\nu}dx^\mu dx^\nu$ is the line element of the flat
metric. Then the field equation reads :
\begin{equation}
\label{Eqchpmat}\Box_{\textrm{flat}} \varphi(x)= \frac{4 \pi
G}{c^2} A'(\varphi)\sum_A m_A \int \delta^{(4)} \left( x^\mu  -
z_A^\mu \left(\tau_A\right)\right) ds,
\end{equation}
where $\tau$ is the proper time along worldlines. Here we notice
that Nordstr\"om's theory, corresponding to $A(\varphi)=\varphi$,
gives a \textit{linear} equation, so that $\varphi$ is simply the
sum of a constant and of the separate contributions of each point
particle. This a great theoretical advantage over GR. The
linearity of the field equation together with the fact that the
SEP holds in Nordstr\"om's theory motivates our study of this
particular theory.
\section{Dynamics of N point-particles; the two body problem}
Let us consider $N$ point particles whose motion is assumed to be
known. We parameterize their positions $z_A^\mu (\tau_A)$ by their
proper time $\tau_A$. Then the linear field equation
Eq.~(\ref{Eqchpmat}) is solved with
\begin{equation}
\label{Solchp} \varphi(x) = 1-\sum_A \frac{G
m_A}{\rho_A^{\textrm{ret}}(x) c^2}
\end{equation}
where
\begin{equation}
\label{Rhoret} \rho_A^{\textrm{ret}}(x)=
-u^{A}_{\nu}(\tau'_A)(x^\nu -z_A^\nu (\tau'_A)),
\end{equation}
is a scalar distance between $x$ and $z_A(\tau'_A)$, where $u$ is
the four-velocity and $\tau'_A$ is the \textit{retarded} proper
time given by the intersection of the A-th particle world line and
the past light cone of x. Following \cite{Poisson}, we refer to it
as the retarded distance. Let us stress that, contrary to GR or to
any theory of type (\ref{Action}) with $A \neq Id$, \textit{the
gravitational field is thus exactly known}.

On the other hand, we now look for the equation of motion of point
particles in a given external field. Equation~(\ref{Conservtilde})
shows that test masses are following geodesics of the physical
metric. However, in the two comparable mass body problem, one
cannot neglect the effect of the proper field on the motion (the
so-called self-force). Actually, Eq.~(\ref{Solchp}) shows that the
self-field is singular on bodies world lines, so that their exact
equation of motion (including the self-force) is not yet defined.

The same problem occurs in classical electromagnetism. Dirac
\cite{Dirac} addressed this issue in EM by using the local
conservation of energy-momentum in a small 3-tube surrounding the
world lines, and derived the correct (known as the Lorentz-Dirac)
equation of motion. In \cite{NotreArticle} we adapt this procedure
to our specific case using Eq.~(\ref{Conservtilde}), following
\cite{Damourthese} and \cite{Poisson}. The equation of motion of
body $A$ in a smooth external field $\varphi^{\textrm{ext}}$
finally reads
\begin{equation}
\label{EqM}
\frac{d}{ds}\left(\left(1+\varphi^{\textrm{ext}}\right)u_\nu
\right)+\frac{G m_A}{3 c^2}
\left(\dot{u}^2u_\nu-\ddot{u}_{\nu}\right)+\partial_{\nu}\varphi^{\textrm{ext}}=0,
\end{equation}
where $m_A$ is the mass of body $A$, $u$ denotes its four-velocity
and the dot means the derivative w.r.t the proper time. As far as
the $N$-body problem is concerned, one should reintroduce the
indices and write the external field with the help of
Eq.~(\ref{Solchp}). The self-forces term in Eq.~(\ref{EqM}) are
those which are proportional to $m_A$ (because neglecting the
self-force means neglecting the mass of the particle), and we see
that they are third derivative of the position, as in the
Lorentz-Dirac equation.

Let us specialize these results to the two body problem. The above
manifestly covariant form of the equation of motion (EOM) can be
written in a more useful way in terms of cartesian coordinates
$(ct,x)$. We write $z_A^\mu (t)=(c t,\bm{z}_A(t))$, and similarly
for $B$. Let $\bm{v}_A(t)=d\bm{z}_A/dt$ be the three-velocity. We
define its velocity as $\bm{\beta}_A(t)=\bm{v}_A(t)/c$, and the
Lorentz factor $\gamma_A$ by $(1-\bm{\beta_A}^2)^{-1/2}$. For any
position of body $A$ at time $t$, there is a unique retarded
position of body $B$ given by the intersection of the past light
cone of $z_A^\mu (t)$ and the world line of $B$. We shall write it
as $z_B^\mu (t_{\textrm{ret}})=(c
t_{\textrm{ret}},\bm{z}_B(t_{\textrm{ret}}))$, where
$t_{\textrm{ret}}$ is the retarded time. The retarded distance is
then
\begin{equation}
\label{Rhoretz} \rho_B^{\textrm{ret}}[z_A(t)]=
-u^{B}_\nu(t_{\textrm{ret}})\left(z_A^\nu (t)-z_B^\nu
(t_{\textrm{ret}})\right),
\end{equation}
and is simply denoted $\rho_B^{\textrm{ret}}$. With the above
notations, it is straightforward to show that the dynamics of body
$A$ at time $t$ is given by :
\begin{equation}
\label{VEOMmamb} c\dot{\bm{\beta}}_A=\frac{1}{\gamma_A^2(1-G
m_B/\rho^{\textrm{ret}}_B c^2)} \left[\frac{-G
m_B}{\left(\rho_B^{ret}\right)^2}\left(\bm{\nabla}_{\bm{z}_A} \rho
_B^{\textrm{ret}}+\frac{\bm{\beta}_A}{c}\dot{\rho}_B^{\textrm{ret}}\right)+G
m_A\left(\frac{\gamma_A^3\,\ddot{\bm{\beta}}_A}{3}
+\gamma_A^4\,\left(\bm{\beta}_A\dot{\bm{\beta}}_A\right)\dot{\bm{\beta}}_A\right)\right],
\end{equation}
where all quantities have to be taken at time $t$. Now the dot
means the derivative w.r.t \textit{time}, $d/dt$. The gradient of
$\rho_B^{\textrm{ret}}$ is the one w.r.t the position of body $A$
(see \cite{NotreArticle} for useful formulaes). The left hand side
is just the acceleration of body $A$, and we recognize in the
first term of the right hand side a Newtonian-like term,
responsible of a force proportional to $m_A m_B$, whereas the
second one is the self-force term, proportional to $m_A^{\,2}$.
The exact equation of motion for the two body problem is thus
known.

In the following, we shall restrict to the case of equal-mass
bodies. The EOM then trivially comes from Eq.~(\ref{VEOMmamb})
with $m_A=m_B=m$. In that case we have a symmetry that ensures the
existence of a Lorentz frame in which the center of mass is always
at rest. In this frame, the three-position vectors w.r.t the
center of mass obey the law $\bm{z}_A(t)=-\bm{z}_B(t)$ for all
time $t$. Furthermore the motion can be shown to be plane (see
\cite{NotreArticle}). As a consequence, the motion can be
characterized by only two variables whose choice is free. In the
following we choose $r[\beta(t)]$ and $\beta(t)$, where $\beta(t)$
is the norm of the velocity of $A$ (or $B$) and $r$ is the norm of
$\bm{z}_A$ (or $B$). The reason of this choice is that the
gravitational radiation (and therefore the non-trivial temporal
evolution of the binary system) is encoded in the variations of
$\beta(t)$. The angular velocity $w$ is easily expressed in terms
of these variables.

It must be stressed that our dynamical equation is not a complete
set of equations. Indeed, we have to write the kinematical
equations that determine the retarded position of $B$. If we
define $\delta t^{\textrm{ret}} \equiv t-t_{\textrm{ret}}$ and
$\psi^{\textrm{ret}}$ as the oriented, positive, retarded angle
between $\vec z_B(t_{\textrm{ret}})$ and $-\vec z_A(t)$, where the
positive orientation is chosen to be the one defined by the
angular velocity vector, these (implicit) equations are
\begin{equation}
\label{EOMdtpsi} c \delta t^{\textrm{ret}}
=\sqrt{r^2[\beta(t)]+2r[\beta(t)]r[\beta(t_{\textrm{ret}})]\cos
\psi^{\textrm{ret}}+r^2[\beta(t_{\textrm{ret}})]},\,
\psi^{\textrm{ret}}=\int_{t-\delta t^{\textrm{ret}}}^{t}
\omega(t') dt'.
\end{equation}
This set of four equations Eq.~(\ref{VEOMmamb}) (with $m_A=m_B=m$)
and Eq.~(\ref{EOMdtpsi}), which we will refer to as the equations
of motion (EOM), are now sufficient to determine the entire
motion. In the next section, we look for a circular motion. This
unphysical motion will be used as an initialization of an
iterative method that will construct order by order the actual
inspiral motion, see Section V.

\section{The circular configuration, its energy and the ICO}
We now restrict ourselves to equal-mass binaries. In order to
obtain a stationary, circular solution, we have to use the time
symmetric Green function when solving the field equation,
\textit{i.e.}, we have to consider as much outgoing waves as
ingoing ones. We show in \cite{NotreArticle} that the circular
solution then reads
\begin{equation}
\label{r0} r_0(\beta)= \frac{G
m}{c^2}\frac{1+2\beta^2-\beta^4\cos\psi_0
-2\left(\beta+\beta^3\right) \sin\left(\frac{\psi_0}{2}\right)}{4
\gamma \beta^2 \cos\left(\frac{\psi_0}{2}\right) \left(1+\beta
\sin \left(\frac{\psi_0}{2}\right)\right)},
\end{equation}
where the radius $r_0$ denotes the distance of one body to the
center of mass, and $\beta=v/c$ is the constant orbital velocity
of the bodies. In the literature the notion of separation is more
often used and is given here by $2 r_0$. Here $m$ denotes the mass
of each body, and the retarded angle $\psi_0$ is given by the
implicit equation
\begin{equation}
\label{PsiCirc} \frac{\psi_0}{2}=\beta \cos
\left(\frac{\psi_0}{2}\right).
\end{equation}
The index ``$0$'' simply refers to the circular case. Note that to
the leading order, the radius reads $r_0 \sim r_{Newton}= G m/4
\beta^2 c^2$ just as in Newton's theory. In order not to show this
leading but trivial part of the function $r_0$, we plot in the
right panel of Fig.~(\ref{FigureFonctions}) the ratio
$r_0/r_{Newton}$ as a function of $\beta$. We see that it goes to
zero if the velocity goes to $1$.

Since Nordstr\"om's theory shares many features of the general
relativity, it is interesting to look for the exact expression of
the energy of the binary being in a circular motion, in order to
investigate in a non-perturbative way the existence or not of the
Innermost Circular Orbit (ICO), which is an important notion in
numerical relativity even if it does not correspond to any
physical observable.

In \cite{NotreArticle} we compute the energy of the binary using
the Fokker action associated to the initial action of the theory.
The result is surprisingly simple since it reads :
\begin{equation} \label{Energie}
E(\beta)=\frac{2 m c^2}{\gamma},
\end{equation}
which have also been found independently by Friedman and
Ury$\bar{u}$ \cite{Friedman} in a related context. The energy of
the binary thus goes to zero in the ultra relativistic limit, so
meaning that all the initial energy of the binary has been carried
away by radiation. Since the limit $v=c$ is also the limit $r_0
=0$, we shall consider that the two bodies have melt each other
This remaining point particle has a vanishing energy, and thus
does not actually exist. The two initial bodies have thus been
entirely evaporated into gravitational radiation. It must however
be stressed that this conclusion would hold only if the physical
motion were a succession of circular orbits of decreasing radius,
which is obviously not the case. Actually the inspiral motion can
be seen as such a succession only if the damping timescale is much
greater than the orbital one, an assumption that breaks down in
the relativistic regime, as we will show in Section VI.

\begin{figure}
\caption{\label{FigurePlotICO} The PN ICO's up to $38$ PN. ICO's
do not necessarily exist at each PN order. The series seem to
converge whereas this can not be the case. ICO's found in GR are
also shown (extracted from \cite{BlanchetISCO3PN}).}
\includegraphics[width=12 cm]{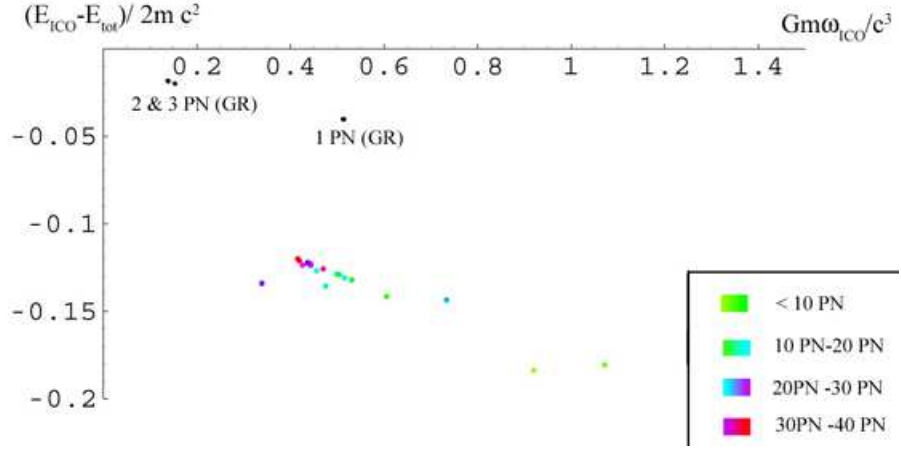}
\end{figure}

The ICO is defined as the separation of the companions for which
the energy is minimum, if such a point exists \cite{BaumgarteICO}.
Now, if the separation decreases and becomes smaller than the ICO,
the motion cannot remain (quasi) circular unless energy of binary
grows, which is impossible because of gravitational radiation.
Passing through the ICO may therefore be the definition of the
beginning of the plunge phase. Since
\begin{equation} \frac{d E}{d r}= \left(\frac{dr_0(\beta)}{d
\beta}\right)^{-1} \frac{d E}{d \beta},
\end{equation}
does not vanish (unless $\beta=0$, that is $r=\infty$), there is
no ICO in Nordstr\"om's theory. However, if one now works in a PN
truncation, one gets e.g. at the second order
\begin{equation}
E^{\textrm{2PN}}(r)=2m c^2\left[1-\frac{1}{8}\left(\frac{G m}{r
c^2}\right)-\frac{5}{128}\left(\frac{G m}{r
c^2}\right)^2+\frac{27}{1024}\left(\frac{G m}{r c^2}\right)^3 +
\mathcal{O}\left(\frac{G m}{r c^2}\right)^4\right].
\end{equation}
This $2$PN energy has a turning point for a separation being
roughly $1.08 G m/c^2$. It thus shows that order-dependent ICOs
can be found although the non-perturbative ICO does not exist. We
find such PN ICOs at 2, 3, 5, 6, 8, 9,\ldots  PN orders. In
Fig.~(\ref{FigurePlotICO}) we have plotted up to very high PN
orders (38 PN) the position of the ICO and placed the points
obtained in general relativity (extracted from
\cite{BlanchetISCO3PN}). In this figure, we plot the angular
frequency of the binary at the ICO as a function of the energy of
the binary at the ICO, normalized to the total energy :
$(E_{\textrm{ICO}}-2 m c^2)/(2mc^2)$ (see \cite{BlanchetISCO3PN}).

The ICOs we find here are quite faraway from those found in GR (by
numerical study or PN expansions). The most striking feature of
this plot is that the PN ICOs seem to converge, whereas there is
no ICOs in the exact theory.

\section{Construction and convergence of a perturbative solution}
Beyond the circular motion, which is not physical unless advanced
and retarded waves are considered, we shall now construct an
inspiral motion, solution of the only retarded EOM. When deriving
a perturbative solution to the EOM, one find convenient to look
for the orbital acceleration $\dot{\beta}$ as a function of
$\beta$
\begin{equation}
\label{Solutionf} \dot \beta(t)=f[\beta(t)],
\end{equation}
which, together with the radius as a function of $\beta$,
completely characterizes the motion. The perturbative expansion of
the EOM goes as follow. We look for function $f$ and $r$ in power
of a parameter $\varepsilon$, as $f = f_0+ \varepsilon f_{1}+
\ldots+ \epsilon^{n} f_{n}+ \ldots$ and similarly for $r$, where
the $f_i$ and $r_i$ are unknown function of $\beta$, and
$\varepsilon$ is an arbitrary parameter. We choose $f_0=0$ in
order to recover an expansion whose 0-th order corresponds to the
circular motion. It is important to note that this choice is
analytically unmotivated, but rather corresponds to an
\textit{initial condition assumption}.

We have proven in \cite{NotreArticle} that this expansion
converges in powers of $\beta^5$, which means that the property
$f_i\sim \mathcal{O}(1)$ and $r_i\sim \mathcal{O}(1)$ is satisfied
when $\varepsilon = \beta^5$. Moreover, this amplitude is directly
related to the first non-vanishing multipolar radiation of the
binary, namely the quadrupolar one : $\dot E/E \sim \beta^5$,
where $E$ is the energy of the binary. Note also that this
expansion is not a PN one since each functions $f_i$ and $r_i$ are
found to possess a complicated series expansion, and are not just
monomials. For instance the 0-th order provides an
\textit{analytical} formula linking the radius of the circular
motion to its orbital velocity, see Eq.~(\ref{r0}). This method
thus gives to each order many non-perturbative information.

\section{Analytical and numerical results}
We have run the previous algorithm up to the second order. It
means that we have derived the exact, analytical expression of
$f_1$, $f_2$, $r_1$ and $r_2$. We do not write explicitly these
functions here since their expression are quite heavy, but a
simple code enables to find them. These functions $f$ and $r$
fully characterize the motion, so that we will refer to these
first and second order as the first and second order motion.

\begin{figure*}[ht]
\label{FigureFonctions}
 \centerline{\includegraphics[width=8cm]{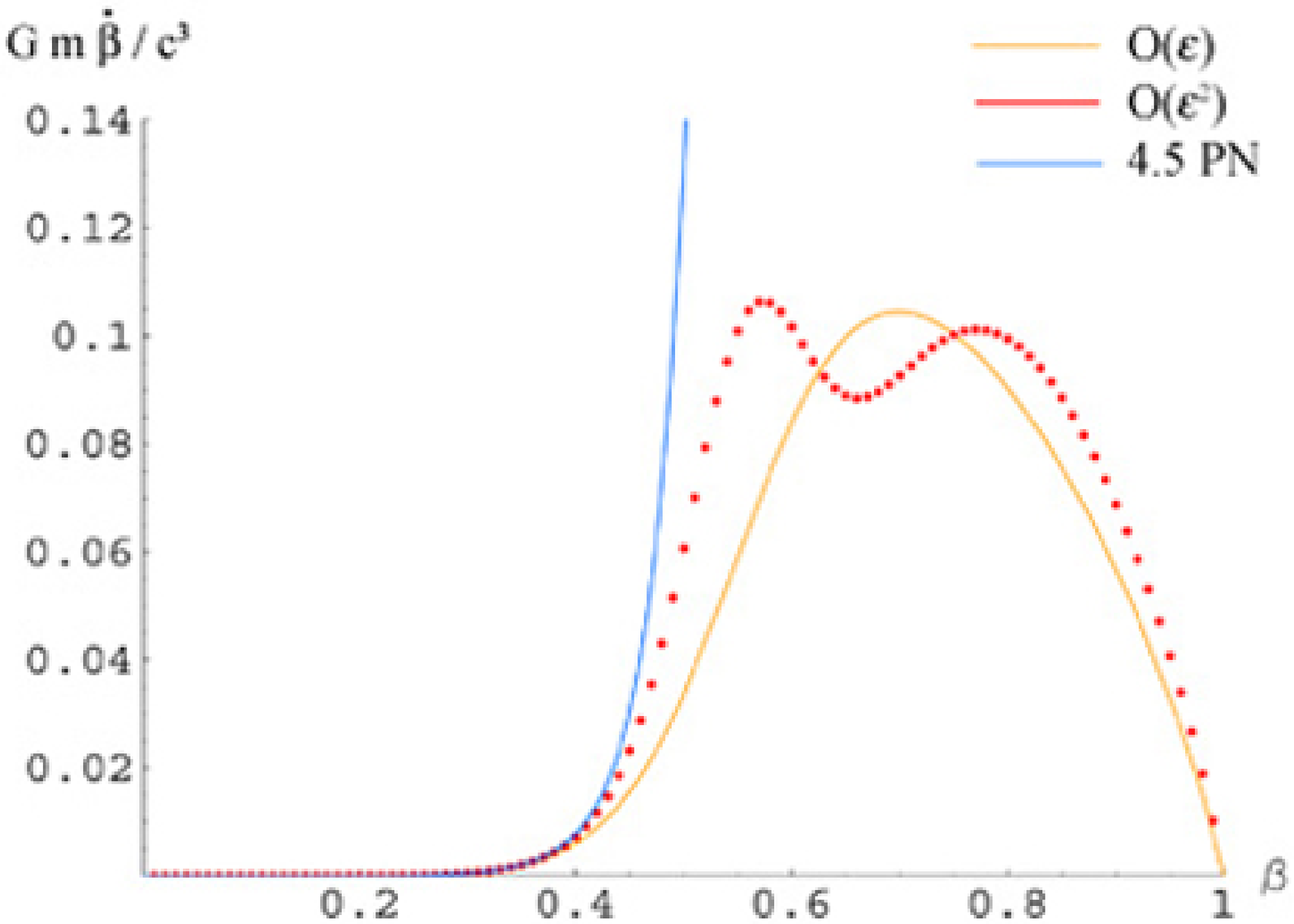}
              \includegraphics[width=8cm]{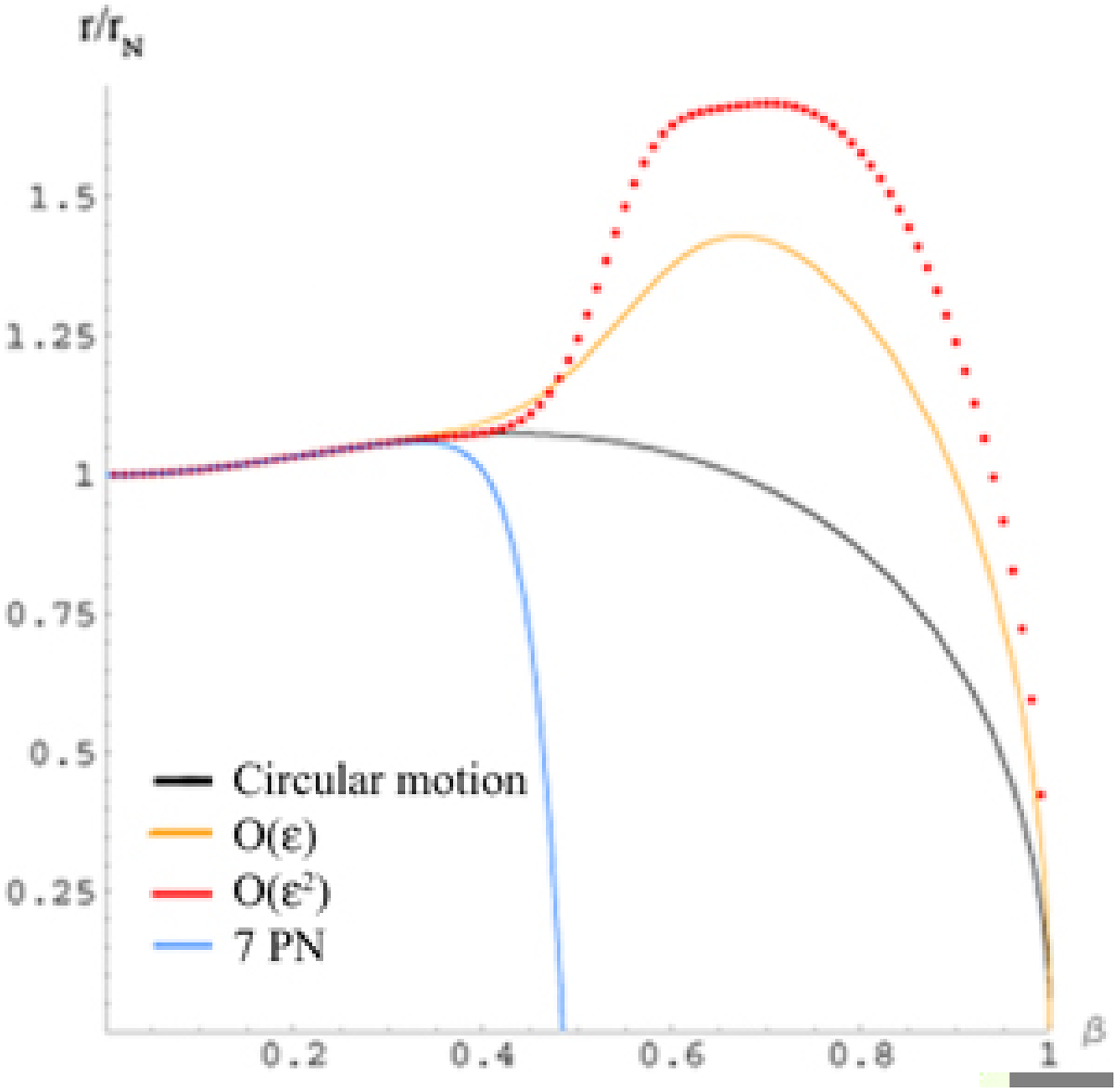}}
  \caption{(left) The orbital acceleration as a function of $\beta$: the first order and second order solutions, compared to the $4.5$ PN one.
  (right) The radius of the orbit as a function of $\beta$. We plot the circular solution, and the first and second order corrections to it.
  The radius at $7$ PN is also given.}
\end{figure*}

We can write the explicit and exact PN expansion of these two
solutions. We show in \cite{NotreArticle} that the first order of
perturbation give both the acceleration and the radius exactly up
to 4.5 PN, whereas the second order provides a correct radius up
to $7$ PN. These results are :
\begin{equation}
\label{DLfinalAcc}\dot \beta_{\textrm{4.5 PN}} = \frac{c^3}{G m}
\left(
\frac{512}{15}\beta^9-\frac{4864}{35}\beta^{11}+\frac{2316736}{2835}\beta^{13}
-\frac{1216928}{297}\beta^{15}
+\frac{5973800428}{289575}\beta^{17}+\mathcal{O}\left(\beta^{19}\right)
\right)
\end{equation}
\begin{equation}
\label{DLfinalR} r_{\textrm{7 PN}} = \frac{G m}{c^2}
\left(\frac{1}{4 \beta^2}+\frac{1}{4}-\frac{5}{4}\beta^2
+\frac{131}{36}\beta^4-\frac{617}{60}\beta^6-\frac{883}{84}\beta^8+\frac{111900409}{170100}\beta^{10}-\frac{119811886267}{13097700}\beta^{12}+\mathcal{O}\left(\beta^{13}\right)\right)
\end{equation}
Let us now focus on the entire non perturbative functions. In the
left panel of Fig.~(\ref{FigureFonctions}), we plot the orbital
acceleration as a function of $\beta$. We show the behavior of
this acceleration for the first and second order motion, and we
also plot the 4.5 PN expansion of it given in
Eq.~(\ref{DLfinalAcc}). We also plot in the right panel of
Fig.~(\ref{FigureFonctions}) the radius as a function of $\beta$,
normalized to the Newtonian radius $r_\textrm{Newton}=Gm/4\beta^2
c^2$. We show the circular radius, the first and second order one,
and the 7PN radius given by Eq.~(\ref{DLfinalR}).

\begin{figure*}[ht]
\label{FigureTemps}
 \centerline{\includegraphics[width=8cm]{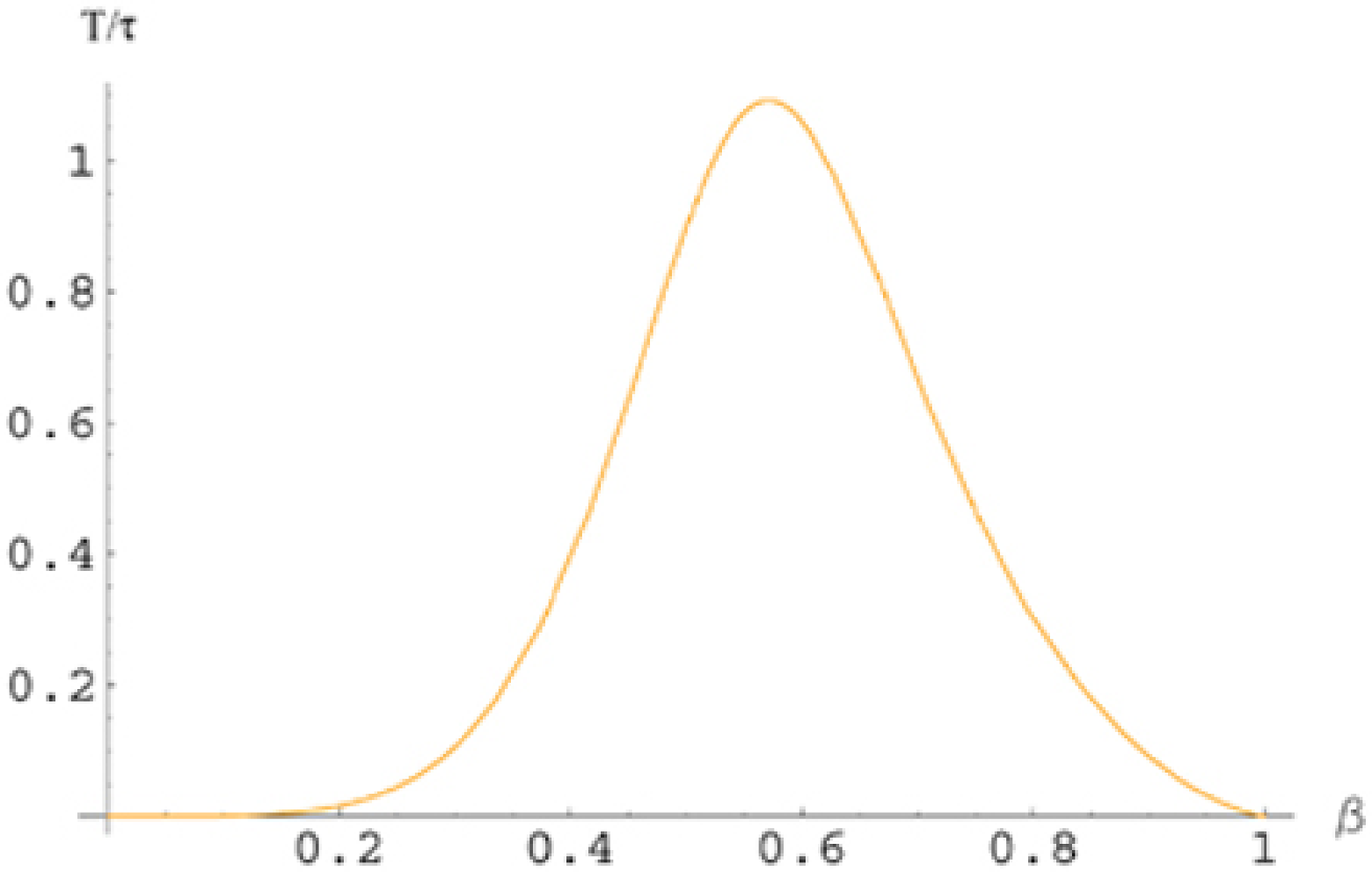}
              \includegraphics[width=8cm]{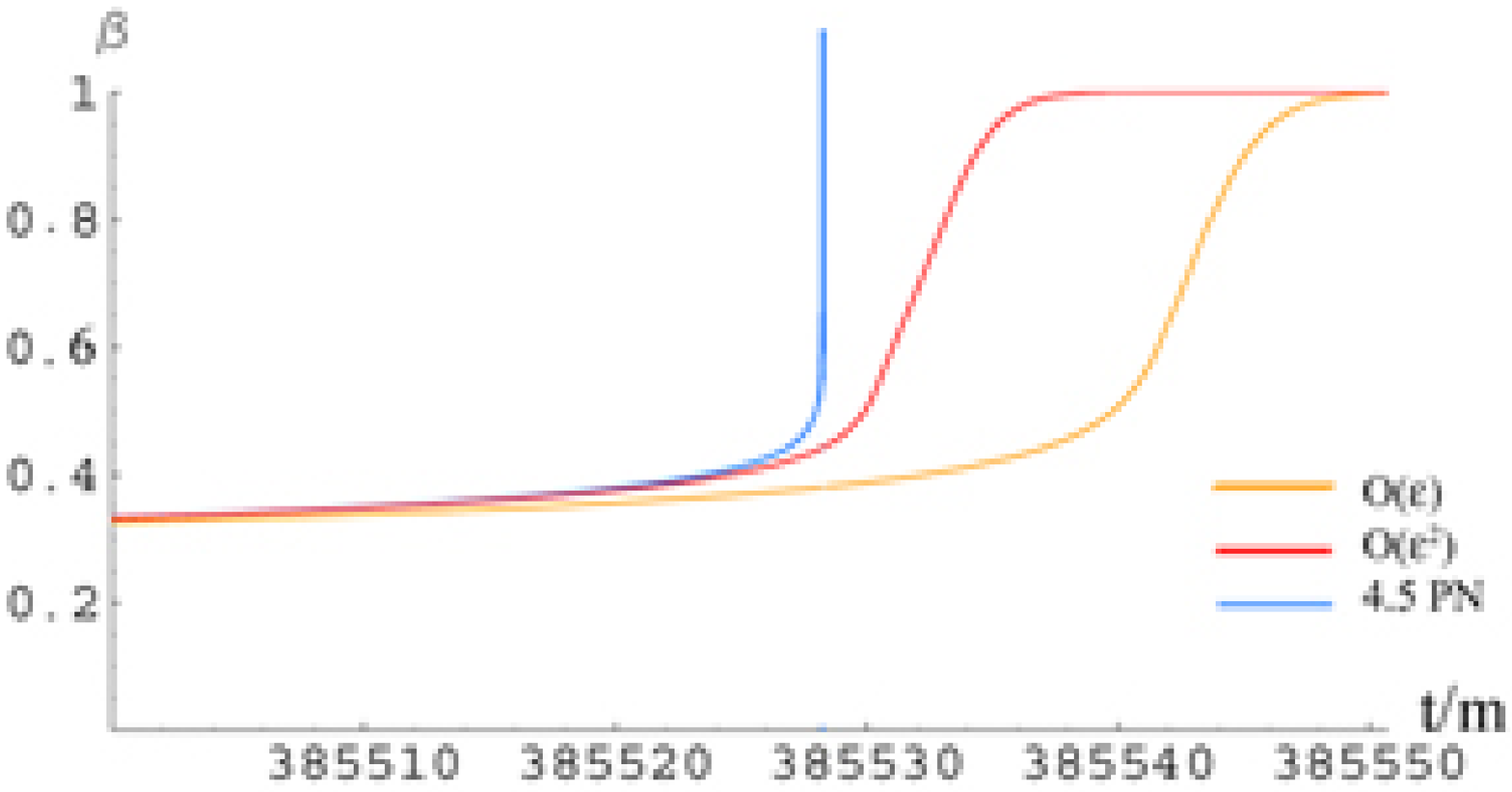}}
  \caption{(left) The ratio of the delay-time over the dynamical damping timescale as a function of $\beta$ at first order.
  These two timescales become comparable in the relativistic regime.(right) The behavior of the velocity in time, using
  the first order solution, the second order one and the PN
solution one.}
\end{figure*}

In \cite{NotreArticle}, we also investigate the so-called
Quasi-Equilibrium (QE) scheme which has been extensively used in
GR (see, e.g, \cite{QEinGR},\cite{QEinGR2}) and discuss its
validity in Nordstr\"om's theory. This approximation holds if the
dynamical (damping) timescale is much greater than the orbital
period, so that the gravitational radiation can be neglected at
first approximation, and the motion is quasi-circular. The left
panel of Fig.~(\ref{FigureTemps}) represents the ratio (at first
order) of the orbital period over the dynamical damping timescale.
We thus see that the QE approximation is an excellent one in the
non-relativistic \textit{and} in the ultra-relativistic regime.
However, in the intermediate regime ($\beta \sim 0.6$) the
dynamical timescale becomes comparable and even shorter than the
orbital period, so that the orbit is highly non-circular and the
approximation underlying the QE-scheme is strongly broken. This is
the plunge phase of the binary.

\begin{figure*}[ht]
\label{FigureSolutionsPlan}
 \centerline{\includegraphics[width=7cm]{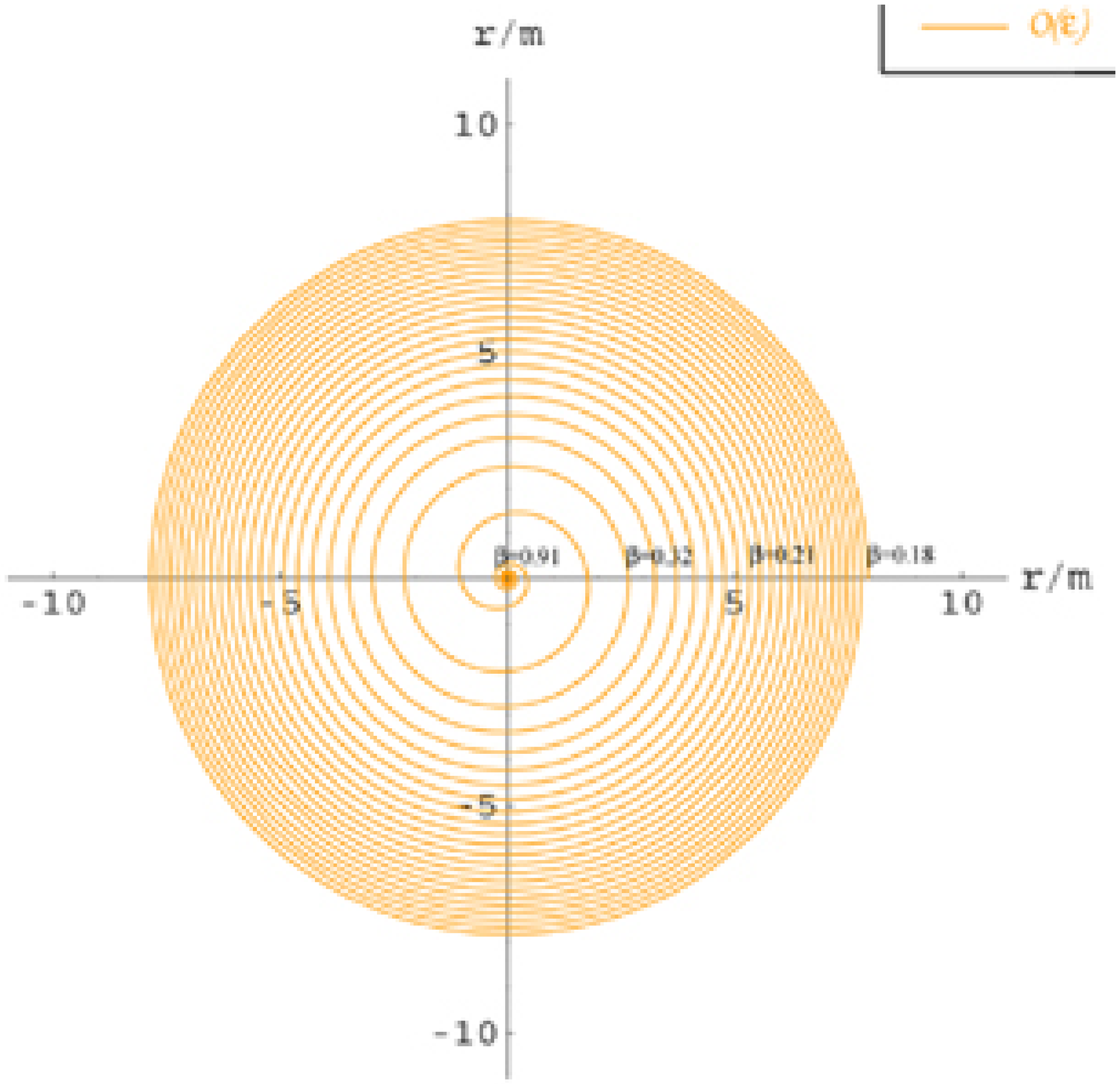}
              \includegraphics[width=7cm]{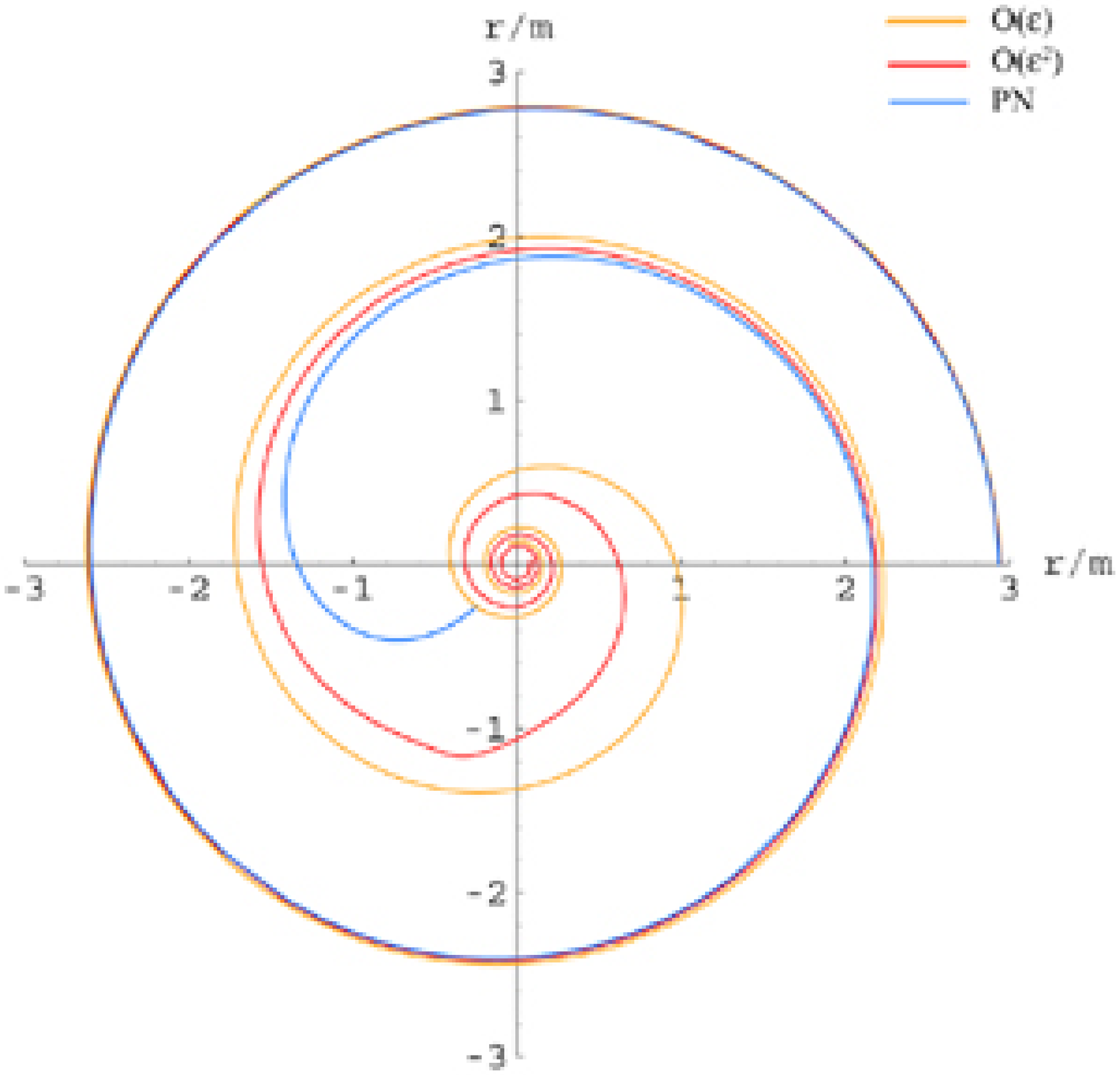}}
  \caption{(left) The motion of one companion, from $\beta=.18$ to $\beta=.91$, using the
  first order result $f_1$. (right) The ultrarelativistic phase and the comparison of the first
  order solution (external line), the second order one (middle line) and the PN solution (interior
  line).}
\end{figure*}

A numerical integration of Eq.~(\ref{Solutionf}) yields the
behavior of the norm of the velocity $\beta$ as a function of
time, plotted in right panel of Fig.~(\ref{FigureTemps}). The
three curves corresponds to the 4.5 PN result, the first order and
second order result, starting at $\beta=0.1$. The three results
are very similar until $\beta \sim 0.4$. A major difference,
however, between the PN solution and the non-perturbative
solutions is that it cannot take into account the pole of special
relativity, so that the PN solution diverges : $\beta(t) \to
\infty$ in a finite time. The non-perturbative solutions, on the
contrary, includes effects of special relativity, notably the fact
that $\beta$ cannot be greater than $1$. It has the major effect
to delay the moment of ``coalescence'' of the two body.
Finally we plot the motion of one of the companion over a large
range of $\beta$, and compare the results of the PN approximation,
the first and second order approximations. This is shown in
Fig.~{\ref{FigureSolutionsPlan}}. Although the behavior of the PN
solution is quite bad near $\beta=1$, we see that the three
solutions agree very well until a separation of order $4 m$ (a
radius $2 m$), which is not too much compared to the Schwarzschild
radius. This might be seen as a good news for PN expansions.
\section{Conclusion and perspectives}
The aim of this work was to extract as much as possible analytical
results concerning the two-body problem in Nordstr\"om's theory,
which is without doubts the purely scalar theory of gravity
closest to GR. Once again, these analytical results are not
\textit{a priori} of direct interest as far as the construction of
relevant templates is concerned, but are interesting since they
enable to estimate the validity of methods used in the two-body
problem in GR. For instance, analytical results derived here
should be useful to test the efficiency of numerical codes. Those
analytical results are already summarized in the introduction.

Of course, an exact solution of the equation of motion may be one
day derived, and that will be of great interest. Further work on
the validity of the QE-scheme may also be done, see
\cite{NotreArticle}. It could also be interesting to examine the
behavior of the PN expansion in the plunge phase, compared to a
numerical solution of the full equation of motion. We could
therefore check if the PN predictions are very bad ones in the
ultra relativistic limit (as expected, since the small parameter
goes to $1$), or if, due to some non-trivial cancellations (note
for example the alternate signs in Eq.~(\ref{DLfinalAcc}) and
Eq.~(\ref{DLfinalR})), the PN picture reveals itself to be a good
one. Such an alternation has already been observed in PN expansion
of GR, see \cite{Alternancesignes}. This behavior has not been
explained yet, and we suggest that it might be understood in
Nortsr\"om's theory.

\bibliographystyle{aipproc}

\end{document}